 \documentstyle[twocolumn,aps]{revtex}
\begin{document}
\preprint{ {\bf /} {\bf /} }
\draft
\title{Extension of Bertrand's theorem and factorization
of the radial Schr\"odinger equation}
\author{ Zuo-Bing Wu and  Jin-Yan Zeng} 
\address{Department of Physics, Peking University,
Beijing 100871, China}
\date{\today}
\maketitle

\begin{abstract}
The Bertrand's theorem is extended, i.e. closed orbits still may exist for  
central potentials other than the power law Coulomb potential and isotropic harmonic oscillator.
It is shown that for the combined potential $V(r)=W(r)+b/r^2$ ($W(r)=ar^{\nu}$), 
when (and only when)
$W(r)$ is the Coulomb potential or isotropic harmonic oscillator, closed orbits
still exist for suitable angular momentum. The correspondence between the closeness of 
classical orbits and  the existence of raising and lowering operators derived from 
the factorization of the radial Schr\"odinger equation is investigated. 
\end{abstract}

\pacs{PACS number(s): 03.65. -w; 03.65. Ge}
\narrowtext
\section{Introduction}

The orbit of a classical particle in a central field, due to the angular momentum 
conservation, must lie in a plane perpendicular to the angular momentum.
However, the orbit is, in general, not closed. In classical mechanics there is
a famous Bertrand's theorem\cite{Bertrand,Goldstein}, which says that {\it the only central forces 
that result in closed
orbits for all bound particles are the inverse square law and Hooke's law}. For
the two central potentials, apart from the energy and angular momentum, there exists
an additional conserved quantity, which implies a higher dynamical symmetry than
the geometrical symmetry (space isotropy).

Over fifty years ago, Schr\"odinger introduced the factorization method\cite{Schrodinger} 
to treat
the eigenvalue problem of a one-dimensional harmonic oscillator and 
the quantized energy eigenvalues of a harmonic oscillator are connected by energy raising 
and lowering operators.  The factorization method was generalized in the 
supersymmetry quantum mechanics (SSQM)\cite{CKS} with the help of the concept of 
supersymmetry and shape invariance. SSQM mainly focuses
on the factorization of one-dimensional Schr\"odinger equation (including
the radial Schr\"odinger equation of a particle in a central field)\cite{HR}
and the relation of eigenstates between two quantum systems (supersymmetric
partner). 

It was shown\cite{LLZ,NG} that only for the Coulomb potential and isotropic harmonic oscillator 
the radial Schr\"odinger equation can be factorized and {\it both the angular momentum
and energy raising and lowering operators can be constructed}.
This reminds us that there may exist a certain connection
between the factorization of radial Schr\"odinger equation and
the closeness of classical orbits.
Careful examination shows that in the derivation of Bertrand's theorem and 
in \cite{LLZ}, a power law central potential ($V(r)=ar^{\nu}$)
was assumed. For such a power law central potential,
the Bertrand's theorem does hold. However, more careful analysis shows that if the restriction
of power law form central potential is relaxed, the Bertrand's theorem 
 need reexamination. In sect. II,
it is shown that for the combined type of central potential 
$V(r)=W(r)+ b/r^2$ ($W(r)=ar^{\nu}$), when $W(r)$ is the Coulomb potential or isotropic 
harmonic potential, there still exist closed orbits for suitable angular momenta. 
The factorization of the corresponding radial Schr\"odinger equation and 
the connection between the closeness of classical orbits and the existence of raising
and lowering operators are investigated in sect. III. 
A brief summary is given in sect. IV. 
 
\section{Extension of Bertrand's Theorem}
\label{sec:bert}

It will be shown that if the restriction of a pure power law central potential is
relaxed, the Bertrand's theorem may be extended. In particular, when an additional
central potential $b/r^2$ is added to $W(r)=ar^{\nu}$,
\begin{equation}
V(r)=W(r)+b/r^2,
\label{eq1.1}
\end{equation}
we will show that when and only when $W(r)$ is the Coulomb potential or isotropic
harmonic potential, there still exist closed orbits for some suitable angular momenta.

The equations of motion of a particle in the central potential (\ref{eq1.1}) is expressed as
($\mu=1$)
\begin{equation}
\begin{array}{l}
\dot{r}=p_r, \dot{\theta}=L/r^2,\\
\dot{p_r}=L^2/r^3+f(r), \dot{L}=0,\\
\end{array}
\label{eq1.2}
\end{equation}
where $L$ is the angular momentum and $f(r)=-dV/dr=g(r)+2b/r^3$, $g(r)=-dW/dr=-a \nu r^{\nu-1}$.
It was shown\cite{Goldstein} that for a circular orbit at radius $r_0$ to be possible,
the force must be attractive and 
\begin{equation}
E=L^2/2r_0^2+V(r_0), ~~~f(r_0)=-L^2/r_0^3. 
\label{eq1.3} 
\end{equation}

To guarantee the stability of a circular orbit $r=r_0$, it is required\cite{Goldstein} that the second derivative
of the effective potential $U(r)=V(r)+L^2/2r^2$ must be positive, i.e.

\begin{equation}
-f'(r_0)-3f(r_0) / r_0>0. 
\label{eq1.4} 
\end{equation}
It is noted that the inclusion of the additional force $2b/r^3$ does not alter the
stability condition (\ref{eq1.4}), so we have

\begin{equation}
-g'(r_0)-3g(r_0) / r_0>0,~~~ {\rm i.e.} ~~~a\nu(\nu+2)>0.
\label{eq1.5} 
\end{equation}
Thus $\nu<-2$ or $\nu>0$ for $a>0$, and  $-2<\nu<0$ for $a<0$.
Since we are interested in the bounded motion,  the case $\nu<-2$ for $a>0$ is excluded.
Thus, we obtain the stability condition for circular orbits: $a \nu >0$, i.e.,
$g(r)<0$. 

To seek the condition for closeness of an orbit, 
let us consider the orbit equation\cite{Goldstein} for potential (\ref{eq1.1})
\begin{equation}
\frac{d^2u}{d \theta^2} +(1+\frac{2b}{L^2}) u=- \frac{1}{L^2} \frac{d}{du} W(\frac{1}{u}),~~~
u=\frac{1}{r}.
\label{eq1.6} 
\end{equation}
Introducing $\kappa=\sqrt{1+2b / L^2}$, $\xi=\kappa u$ and $\phi=\kappa \theta$,
we get
\begin{equation}
\frac{d^2\xi}{d \phi^2} +\xi=J(\xi),~~~J(\xi) \equiv- \frac{1}{L^2} \frac{d}{d \xi} W(\frac{\kappa}{\xi}). 
\label{eq1.7} 
\end{equation}
Using (\ref{eq1.3}), it can be shown that at $\xi=\xi_0=\kappa/r_0$, $J(\xi_0)=\xi_0$.
In addition, using (\ref{eq1.5}) and $g(r)<0$, one get 
$\beta^2 \equiv 1-J'(\xi_0) =3+\frac{r}{g}\frac{dg}{dr}|_{r=r_0}>0$, then $\beta^2=\nu+2$.
Thus, the stability condition for circular orbits also may be 
 expressed as: $\beta^2>2$ for $a>0$, and $2>\beta^2>0$ for $a<0$. 

Using the Taylor expansion of $J(\xi)$ at $\xi_0$, a deviation of an orbit from a circle
 $\xi_0$, $\delta \xi=\xi-\xi_0$, is governed by
\begin{equation}
\begin{array}{l}
 \frac{d^2\delta \xi}{d \phi^2} + \delta \xi= J'(\xi_0) \delta \xi+ 
\frac{1}{2} J''(\xi_0) (\delta \xi)^2 \\
+\frac{1}{6} J'''(\xi_0) (\delta \xi)^3 +O((\delta \xi)^4).
\end{array} 
\label{eq1.77} 
\end{equation} 
 
For a small deviation of an orbit from circularity, to the first order of $\delta \xi$,
(\ref{eq1.77}) is reduced to
\begin{equation}
\frac{d^2\delta \xi}{d \phi^2} +\beta^2 \delta \xi=0. 
\label{eq1.78} 
\end{equation}
The solution to (\ref{eq1.78}) can be expressed as:

\begin{equation}
\delta \xi=a_1\cos( \beta \phi),~~~{\rm i.e.}~~~ \delta u=\frac{a_1}{\kappa}\cos(\beta \kappa \theta). 
\label{eq1.8} 
\end{equation}
It can be seen that for $\beta \kappa=q/p$ ( a rational number, $q$ and $p$ are integers),
when $\theta=\theta_0+2\pi p$, $\delta u=\delta u_0$, i.e., the orbit remains closed.
Thus, we get the condition for closeness of orbits against a small deviation from circularity: 
$\beta\kappa$  {\it must be a rational number}. 

To seek the condition for a closed orbit when the deviation from circularity is considerable, 
$\delta \xi$ can be expressed as the Fourier expansion, i.e., in addition to the fundamental term
in (\ref{eq1.8}), more terms in the Fourier expansion  
should be included in $\delta \xi$. Substituting $\delta \xi$ in (\ref{eq1.77}) and 
using the arguments similar to \cite{Goldstein}, 
 one get $\beta^2$=0, 1 and 4 for closed orbits. Thus, 
we obtain the conclusion: when (and only when) $\beta^2=1$ ($\nu=-1$) for $a<0$,
and $\beta^2=4$ ($\nu=2$) for $a>0$, the orbit is closed. Then,  
$\kappa$ {\it must be a rational number}.
Therefore, for the potential (\ref{eq1.1}), when (and only when)
$W(r)$ is the Coulomb potential or isotropic harmonic oscillator, closed orbits 
still exist for suitable angular momentum $L=\sqrt{2b / (\kappa^2-1)}$, where $\kappa$
is a rational number.

$Example:$ For the valence electron in an alkalis atom $V(r)$ may be approximately expressed as
($e=\mu=\hbar=1$)\cite{Zeng} 
\begin{equation}
V(r)=-\frac{1}{r}-\frac{\lambda}{r^2}, (0<\lambda \ll1). 
\label{eq1.9} 
\end{equation}
In this case, the orbit equation (\ref{eq1.6}) is reduced to
\begin{equation}
d \theta=-du / \sqrt{2E / L^2+2u / L^2-\kappa^2u^2}, 
\label{eq1.10} 
\end{equation}
where $\kappa=\sqrt{1-2\lambda / L^2}$. Integrating (\ref{eq1.10}), we get

\begin{equation}
u=\frac{1}{r}=\frac{1}{L^2\kappa^2}[1+\sqrt{1+2EL^2\kappa^2}\cos\kappa (\theta-\theta_0)], 
\label{eq1.12}
\end{equation}
where $\sqrt{1+2EL^2\kappa^2}=\sqrt{1+2E(L^2-2\lambda)}\geq0$ to make the solution
(\ref{eq1.12}) having sense. For $1+2EL^2\kappa^2=0$, the orbits are 
circular. Because $\kappa < 1$ for $\lambda>0$, no elliptic orbits exist.
However, when $\kappa$ is a rational number, closed orbits (other than ellipses)
still exist. Some illustrative examples are displayed in Fig. 1. 
In Fig. 1(a), $\kappa=\frac{1}{2}(L=\frac{2}{3}\sqrt{6\lambda})$, 
  Fig. 1(b), $\kappa=\frac{2}{3}(L=\frac{3}{5}\sqrt{10\lambda})$ 
and Fig. 1(c), $\kappa=\frac{3}{4}(L=\frac{4}{7}\sqrt{14\lambda})$,
there exist closed orbits around the center with period 2, 3 and 4, respectively. 
 The geometry of closed orbits is 
dependent on $\kappa$ ($L$), but irrelevant to the energy $E$.

\begin{figure}
\caption{The closed orbits of the valence electron 
in an alkalis atom ($\lambda=0.2$). }
\label{fig1}
\end{figure}

\section{factorization of the radial Schr\"odinger equation}
\label{sec:fact}
For a particle in the central field $V(r)$, the energy eigenfunction may be 
chosen as the simultaneous eigenstate of the complete set of conserved observables
($H$,$\hat{\bf l}^2$,$l_z$), and the radial wave function $R_{n,l}(r)=\chi_{n,l}(r) /r$, 
($l$ are nonnegative integers) satisfies the radial Schr\"odinger equation ($\hbar=\mu=1$)

\begin{equation}
\begin{array}{l}
 H \chi_{n,l}(r)  = E_n \chi_{n,l}(r), \\
 H=-\frac{1}{2} d^2 / dr^2+l(l+1) / 2r^2+V(r).
 \end{array}
 \label{eq2.2} 
 \end{equation}
For the potential (\ref{eq1.1}), (\ref{eq2.2}) is reduced to 
\begin{equation}
\begin{array}{l}
[\frac{d^2}{dr^2} -\frac{l(l+1)}{r^2}-2W(r) -\frac{2b}{r^2}] \chi_{n,l}(r) \\
=-2E_{n} \chi_{n,l}(r).
\end{array} 
\label{eq2.3} 
\end{equation}
Let $l'(l'+1)=l(l+1)+2b$, i.e., 
$l'=-\frac{1}{2}+(l+\frac{1}{2}) \sqrt{1+2b / (l+1 / 2)^2}$,
(\ref{eq2.3}) can be rewritten as 

\begin{equation}
\begin{array}{l}
{\cal D}_{l'}(r) \chi_{n,l'}(r) =-2E_{n} \chi_{n,l'}(r),\\
{\cal D}_{l'}(r)= d^2 / dr^2 -l'(l'+1) / r^2-2W(r), 
\end{array}
\label{eq2.4} 
\end{equation}
which has the same form as the usual radial Schr\"odinger equation with the central potential $W(r)$,
but the nonnegative integer quantum number $l$ is replaced by a non-integer $l'$. 
It has been shown\cite{LLZ,NG} that if the power law central potential $W(r)$ is assumed, 
only for the Coulomb potential or isotropic harmonic oscillator the radial Schr\"odinger equation 
can be factorized to derive the raising and lowering operators of $l'$ ($\Delta l'=\pm 1$). 
However, for $\Delta l'=\pm 1$, the shift in $l$ is by no means $\pm 1$. 
Therefore, though the radial Schr\"odinger equation can be factorized for the central
potential (\ref{eq1.1}), no angular momentum raising and lowering operators ($\Delta l=\pm 1$)
can be constructed, which is intimately connected with the breaking of the dynamical symmetry
of Coulomb potential and isotropic harmonic oscillator due to the added term
$b/r^2$ in (\ref{eq1.1}).

For the Coulomb potential and isotropic harmonic oscillator, whose exact eigenvalues
are well known, the energy raising and
lowering operators have be derived by using different approaches in \cite{LLZ,who}. 
For a particle in a more general central potential, the exact energy
eigenvalues are usually unknown. To derive the energy raising and lowering operators, 
we may use the WKB approximation. According to the WKB approximation, the
 eigenvalues of bounded states of (\ref{eq2.4}) take the form\cite{QR} 
 (note: here the usual angular momentum $l$ has been replaced by $l'$):

\begin{equation}
 E_n =\alpha_{\nu} n^{\frac{2\nu}{\nu+2}},
 \label{eq2.5} 
 \end{equation}
where $\alpha_{\nu}=a^{\frac{2}{\nu+2}}2^{\frac{\nu}{\nu+2}}[{\nu\sqrt{\pi}
\frac{\Gamma(\frac{1}{\nu}+\frac{3}{2})}{\Gamma(\frac{1}{\nu})}}]^{\frac{2\nu}{\nu+2}}>0$,
$n=n_r+\frac{l'}{2}+\frac{3}{4}$ for $a>0$, and
 $\alpha_{\nu}=-(-a)^{\frac{2}{\nu+2}}2^{\frac{\nu}{\nu+2}}$
 $[{(-\nu)a\sqrt{\pi}
 \frac{\Gamma(1-\frac{1}{\nu})}{\Gamma(-\frac{1}{2}-\frac{1}{\nu})}}]^{\frac{2\nu}{\nu+2}}$
 $<0$, $n=n_r+\frac{2l'+\nu+3}{2(\nu+2)}$ for $a<0$.
Thus, (\ref{eq2.4}) may be recast into

\begin{equation}
\begin{array}{l}
{\cal D}_n(r) \chi_{n,l'}(r)  = l'(l'+1) \chi_{n,l'}(r), \\
{\cal D}_n(r)=r^2 d^2 / dr^2-2ar^{\nu+2}+2\alpha_{\nu} n^{ 2\nu /(\nu+2)}r^2.\\
 \label{eq2.6} 
\end{array}
 \end{equation}
(1) For $a>0$, the factorizability of (\ref{eq2.6}) requires 
\begin{equation}
\begin{array}{l}
 (r d / dr-\sqrt{2a}r^{\frac{\nu}{2}+1}+A)(r d / dr+\sqrt{2a}r^{\frac{\nu}{2}+1}\\
 +B) \chi_{n,l'}(r)={\cal D}_n(r) \chi_{n,l'}(r)  +AB \chi_{n,l'}(r), 
\end{array}
 \label{eq2.7} 
 \end{equation}
then we have 

\begin{equation}
\begin{array}{l}
 A+B+1=0,\\
 \sqrt{2a}r^{\frac{\nu}{2}+1}(A-B+\nu /2+1)=2\alpha_{\nu}n^{2\nu /(\nu+2)}r^2.\\
\end{array}
 \label{eq2.8} 
 \end{equation}
The solution to (\ref{eq2.8}) is $\nu=2$, $\alpha_2=2\sqrt{2a}$, $A=2n-3 /2$, and 
$B=-(2n-1 /2)$. Thus, (\ref{eq2.7}) is reduced to
\begin{equation}
\begin{array}{l}
[r d /dr-\sqrt{2a}r^2+2n-3 /2][r d /dr+\sqrt{2a}r^2\\
-(2n-1 /2)] \chi_{n,l'}(r)={\cal D}_n(r) \chi_{n,l'}(r)\\
-(2n-3 /2)(2n-1 /2) \chi_{n,l'}(r).
\end{array}
 \label{eq2.9} 
 \end{equation}
 
Similarly, (\ref{eq2.6}) also can be factorized as 
\begin{equation}
\begin{array}{l}
[r d / dr+\sqrt{2a}r^2-(2n+3 /2)][r d / dr-\sqrt{2a}r^2\\
+(2n+ 1 /2)] \chi_{n,l'}(r)={\cal D}_n(r) \chi_{n,l'}(r)\\
-(2n+1 /2)(2n+3 /2) \chi_{n,l'}(r).
\end{array}
 \label{eq2.10} 
 \end{equation} 

Using (\ref{eq2.9}) and (\ref{eq2.10}), one may obtain the energy raising and lowering 
operators
\begin{equation}
\begin{array}{l}
\chi_{n+2,l'}(r)\sim [r d/ dr-\sqrt{2a}r^2+(2n+1 /2)] \chi_{n,l'}(r),\\
\chi_{n-2,l'}(r)\sim [r d/ dr+\sqrt{2a}r^2-(2n-1 /2)] \chi_{n,l'}(r).\\
\end{array}
 \label{eq2.11}
 \end{equation}

(2) For $a<0$, the factorizability of (\ref{eq2.6}) requires  
\begin{equation}
\begin{array}{l}
 (r d /dr-\sqrt{-2\alpha_{\nu}}n^{\nu / (\nu+2)}r+A)(r d /dr\\
+\sqrt{-2\alpha_{\nu}}n^{\nu / (\nu+2)}r+B)\chi_{n,l'}(r)={\cal D}_n(r) \chi_{n,l'}(r)\\
+AB \chi_{n,l'}(r), 
 \end{array} 
 \label{eq2.12} 
 \end{equation}
then we have 

\begin{equation}
\begin{array}{l}
 A+B+1=0,\\
 \sqrt{-2\alpha_{\nu}}n^{\nu / (\nu+2)}r(A-B+1)=-2ar^{\nu+2}.\\
\end{array}
 \label{eq2.13} 
 \end{equation}
The solution to (\ref{eq2.13}) is $\nu=-1$, $\alpha_{-1}=-a^2 /2$, $A=n-1$, and $B=-n$.
Thus, (\ref{eq2.12}) is reduced to
\begin{equation}
\begin{array}{l}
 [r d /dr+ar/n +n-1][r d /dr-ar/n \\
 -n]\chi_{n,l'}(r)={\cal D}_n(r) \chi_{n,l'}(r)  -(n-1)n \chi_{n,l'}(r). 
 \end{array} 
 \label{eq2.14} 
 \end{equation}

Similarly, (\ref{eq2.6}) also can be factorized as 
\begin{equation}
\begin{array}{l}
[r d /dr-ar/n-(n+1)][r d /dr+ar/n \\
+n] \chi_{n,l'}(r)={\cal D}_n(r) \chi_{n,l'}(r)  -n(n+1) \chi_{n,l'}(r).
\end{array}
 \label{eq2.15} 
 \end{equation}
 
Using (\ref{eq2.14}) and (\ref{eq2.15}) and introducing the scaling operator ${\cal M}(k)$ 
defined by ${\cal M}(k)f(r)=f(kr)$, one obtains the energy raising and lowering 
operators
\begin{equation}
\begin{array}{l}
\chi_{n+1,l'}(r) \sim {\cal M}(\frac{n}{n+1}) (r d /dr+ar/n +n) \chi_{n,l'}(r)\\
=[r d / dr+ar / (n+1)+n] {\cal M}(\frac{n}{n+1}) \chi_{n,l'}(r),\\
\\
\chi_{n-1,l'}(r) \sim {\cal M}(\frac{n}{n-1}) (r d /dr-ar/n -n) \chi_{n,l'}(r)\\
=[r d / dr-ar / (n-1)-n] {\cal M}(\frac{n}{n-1}) \chi_{n,l'}(r).\\
\end{array}
 \label{eq2.16} 
 \end{equation}
Thus, we arrive at the conclusion that, for the potential (\ref{eq1.1}), 
only when $W(r)$ is the Coulomb potential
 or isotropic harmonic oscillator,  
 the radial Schr\"odinger equation can be factorized to 
 derive the energy raising and lowering operators. 

For a classical particle in the Coulomb potential $W(r)=-1/r$, the orbit is
always closed for any negative energy ($E<0$) and positive angular momentum $L$,
i.e., an ellipse, of which the length of semi-major axis is $a=1/2|E|$
and the eccentricity $e=\sqrt{1-2|E|L^2}$. This is guaranteed by the existence
of an additional conserved quantity---the Runge-Lenz vector ${{\bf R=p \times L -r}/r}$,
which determines the direction of semi-major axis and the eccentricity
$e=|R|$. In quantum mechanics, the angular momentum and bound energy are quantized.
Therefore, it is not surprising that for the Coulomb potential the radial Schr\"odinger
equation can be factorized and there exist four kinds of raising and lowering
operators connecting various eigenstates of energy and angular momentum\cite{LLZ}.

However, for the combined central potential (\ref{eq1.9}), the dynamical
symmetry $O_4$ is broken and ${\bf R}$ no longer remains constant, and the orbit,
in general, is not closed. It has been shown in sect. II that for suitable angular momenta
$L=\sqrt{2\lambda /(1-\kappa^2)}$ ($\kappa$ is a rational number), the orbits
are still closed (but not ellipses). Therefore, it is understandable 
why no angular momentum raising and lowering operators ($\Delta l= \pm 1$) exist,
but energy raising and lowering operators for a fixed $l'$ (see (\ref{eq2.16})) 
still can be constructed. In fact, for $\lambda=0$, (\ref{eq2.16}) is just the
raising and lowering operators $B(l,n \uparrow)$ and $B(l,n \downarrow)$ given in eq. (33) of \cite{LLZ}.
It is noted that the angular momentum raising and lowering operators $A$, $C$ and $D$ in \cite{LLZ}
no longer exist for the combined potential (\ref{eq1.9}).
For the isotropic harmonic oscillator, the situation is similar. 

\section{Summary}
\label{sec:sum}

The Bertrand's theorem is extended in this paper. For a classical particle in 
a power law central potential $W(r)$, the Bertrand's theorem does hold, i.e.,
only for the Coulomb potential or isotropic harmonic oscillator, closed orbits always
exist for {\it continuous} energy and angular momentum. Accordingly, in quantum 
mechanics there exist both  energy and angular momentum raising and lowering operators
connecting neighboring {\it discrete} energy and angular momentum eigenstates. 
For the combined potential (\ref{eq1.1}), it is shown that when $W(r)$ is the Coulomb
potential or isotropic harmonic oscillator, classical closed orbits (other than ellipse)
still exist for continuous energy, but only for suitable angular momenta. From this one
can understand why in this case there exist only energy raising and lowering operators
in quantum mechanics, but no angular momentum raising and lowering operators. 

This work was supported in part by the Post-Doctoral Foundation of
China (ZBW) and the National Natural Science Foundation of China (JYZ). 



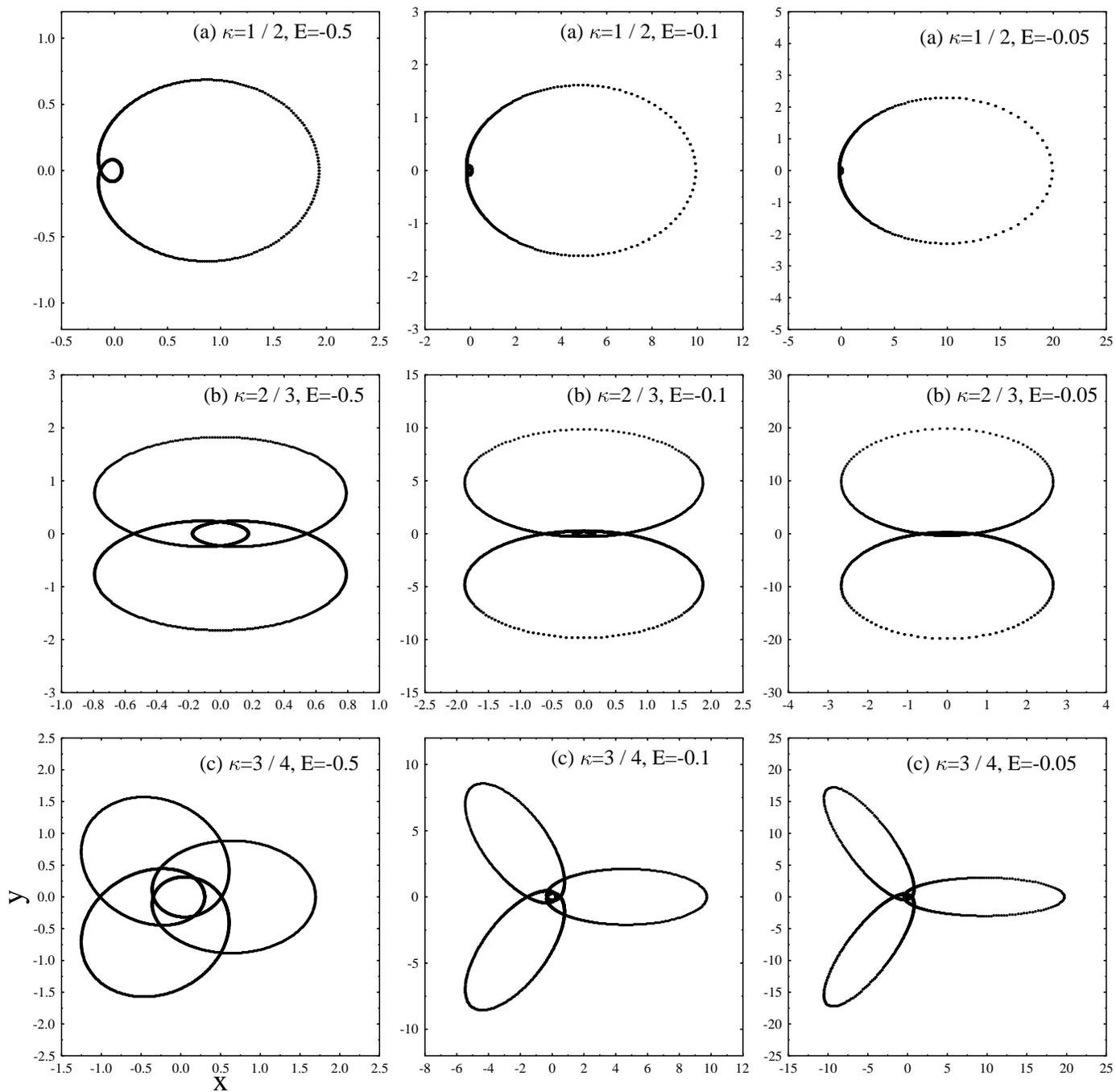
\end{document}